\documentclass[aps,longbibliography,twocolumn,prx,amsmath,caption,amssymb,superscriptaddress,showpacs]{revtex4-1}

\usepackage{graphicx}
\usepackage{color,soul}
\usepackage{mathrsfs}

\newcommand{\dt}[1]{\frac{\partial #1}{\partial t}}
\newcommand{\dx}[1]{\frac{\partial #1}{\partial x}}

\newcommand{\cP}{{\cal P}}
\newcommand{\cV}{{\cal V}}
\newcommand{\A}{[Osm]}
\newcommand{\W}{[Water]}

\renewcommand{\S}{[Waste]}
\newcommand{\Wi}{[Water]^{\mathrm{in}}}
\newcommand{\Ai}{[Osm]^{\mathrm{in}}}
\newcommand{\Si}{[Waste]^{\mathrm{in}}}
\newcommand{\So}{[Waste]^{\mathrm{out}}}
\newcommand{\vW}{\overline{v}_{\rm Water}}

\newcommand{\qO}{{q_{Osm}^{\mathrm{\rm in}}}}
\newcommand{\qS}{{q_{Waste}^{\mathrm{\rm in}}}}
\newcommand{\qWi}{{q_{Water}^{\mathrm{\rm in}}}}
\newcommand{\qWo}{{q_{Water}^{\mathrm{\rm out}}}}
\newcommand{\sop}[1]{#1}

\begin{document}

\title{{Active osmotic exchanger for efficient nanofiltration inspired by the kidney}}

\author{Sophie Marbach}
\author{Lyd\'eric Bocquet}
\affiliation{Laboratoire de Physique Statistique, UMR CNRS 8550, Ecole Normale Sup\'erieure, PSL Research University, 24 rue Lhomond, 75005 Paris, France}
\email{lyderic.bocquet@lps.ens.fr}
\date{\today}

\begin{abstract}
In this paper we investigate the physical mechanisms underlying one of the most efficient filtration devices: the kidney. 
Building on a minimal model of the Henle Loop -- the central part of the kidney filtration --, we investigate theoretically the detailed out-of-equilibrium fluxes in this separation process in order to obtain absolute theoretical bounds for its efficiency in terms of separation ability and energy consumption. 
\sop{We demonstrate that this separation process operates at a remarkably small energy cost as compared to traditional sieving processes, while working at much smaller pressures.  
This unique energetic efficiency originates in the double-loop geometry of the nephron, which operates as an active osmotic exchanger.  
The principles for an artificial kidney inspired filtration device could  be readily mimicked based on existing soft technologies to build compact and low-energy artificial dialytic devices. 
Such a "kidney on a chip" also point to new avenues for advanced water recycling, targeting in particular sea-water pretreatment for decontamination and hardness reduction.}
\end{abstract}
\maketitle



\section*{Introduction}

Most modern processes for water recycling are based on sieving principles: a membrane with specific pore properties allows to separate the permeating components from the retentate~\cite{Elimelech2011}. Selectivity requires small and properly decorated pores at the scale of the targeted molecules, and this inevitably impedes the flux and transport, making separation processes costly in terms of energy. It also raises structural challenges since high pressures are usually required to bypass the osmotic pressure.  Lately nanoscale materials, like state-of-the-art graphene, graphene oxides or advanced membranes~\cite{Karnik2011,Aluru2015,Bakajin2006,Geim2014,Siria2013}, 
have raised hopes to boost the efficiency of separation processes. 
Yet a necessary step for progress requires out-of-the-box ideas operating beyond traditional sieving separation principles. 

In this context it is interesting to investigate how biological systems are able to defy these constraints in their water cycle.  
They often rely on various forms of osmotically driven transport. 
For example in plants, osmosis is harnessed to drive water and sugars over long distances \cite{Bohr,Henton}.
As we discuss in this work, filtration processes can also benefit from osmotic transport: this is the case of the kidney.

Per day, the human kidney is capable of recycling about 200 L of water and 1.5 kg of salt, separating urea from water and salt at the low cost of 0.5 kJ/L ~\cite{Greger1996} while readsorbing $\approx$ 99\% of the water input. 
The core of the kidney separation process lies in the 
millions of parallel filtration substructures called nephrons~\cite{Greger1996}. A striking feature is that the nephrons of all mammals present a precise loop geometry, the so-called Loop of Henle. This loop plays a key role in the urinary concentrating mechanism and has been extensively studied from a biological and physiological point of view\cite{Greger1996,Gottschalk1958,Baylis1989,Stephenson1972,Palatt1973,Thomas2000,Foster1976,Layton2010,Edwards} (see also Supplemental Section 1 for a short review of physiological literature). The nephron operates the separation of urea from water near the thermodynamic limit, $\sim$0.2 kJ/L \sop{(see Supplemental Material Sec. V)}, yet standard dialytic filtration systems, which are based on reverse osmosis and passive equilibration with a dialysate, require more than two orders of magnitude more energy~\cite{Phoenix2009}. 
Some attemps to build artificial devices mimicking the nephron were reported in the literature,
but they rely on biological tissues or cell mediated transport, and cannot be easily scaled up and transferred to other separation devices~\cite{Borenstein2007,Kim2011, Armignaco2015}. Mimicking the separation process occuring in the kidney remains a challenge.

In this work our goal is to take a physical perspective on the separation process at work in the nephron in order to decipher the elementary processes at work.
This allows to propose a simple biomimetic design for an \textit{osmotic exchanger} inspired by the kidney's Loop of Henle -- see Fig.~\ref{fig1} --, 
combining a passive water permeation and an active salt pumping. Such an artificial counterpart can be implemented based on microfluidic elementary building blocks.  

A key feature of the process is merely geometrical: the $U$-shaped loop of Henle is designed as an active osmotic exchanger, with the waste separated from water and salt via a symbiotic reabsorption. Starting wih physiological models of the nephron~\cite{Foster1976,Stephenson1972}, we revisit the detailed out-of-equilibrium flux balance along the exchanger. Our analysis allows us to obtain, to our knowledge for the first time, 
 absolute theoretical bounds for its efficiency in terms of separation. Furthermore, we are able to predict analytically the energetic performance of the separation process. Comparing  to alternative sieving strategies 
 like reverse osmosis and nanofiltration, we show
 that this osmotic exchanger  operates at a remarkably small energy cost, typically one order of magnitude smaller than  other traditional sieving processes, while working at much smaller pressures. 
 Our predictions further assess the key role played by the second part of the loop in order to reach optimal efficiency and low energy cost.

\section{Transport equations in an osmotic exchanger}

\subsection{Geometry of an osmotic exchanger} 
\begin{figure}[h!]
\center
\includegraphics[width = 8.7 cm]{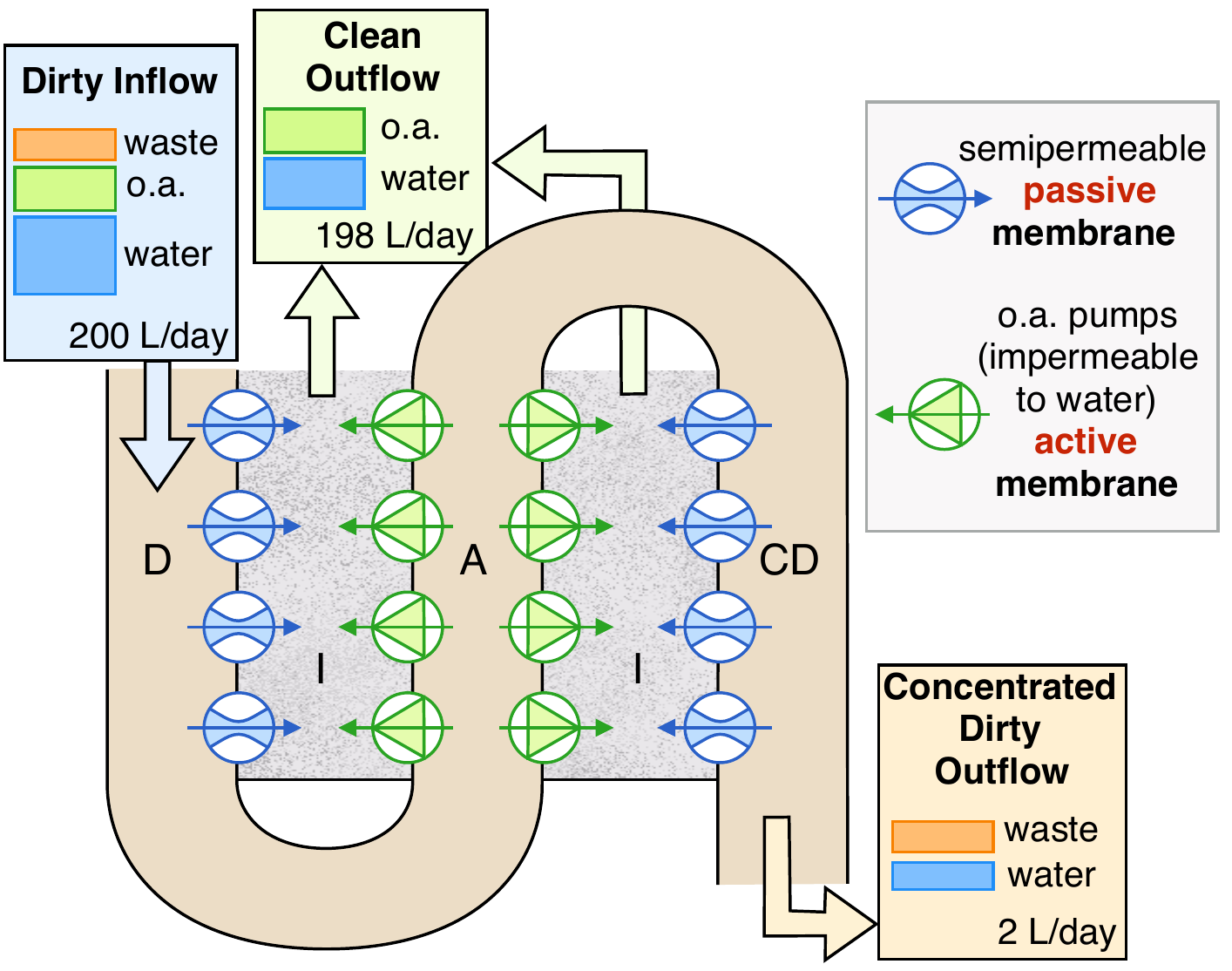}
\caption{{\bf The osmotic exchanger filtration system.} Model for a concentrating device based on the geometry of Henle's Loop in the kidney. Water, waste and salt are carried through the loop of  Henle, constituted of the descending limb (D), ascending limb (A) and continued by the collecting duct (CD). Each limb wall allows exchanges with the interstitium (I), enabling water and salt to evacuate the loop. The remaining waste is concentrated and evacuated by the collecting duct (CD). }
\label{fig1}
\end{figure}

The model we consider is sketched in Fig.~\ref{fig1}. It is a concentrating system that possesses the same primary features as the mammalian nephron: (i) a serpentine geometry consisting \sop{of} (ii) a first U-loop (Henle's Loop, HL) with  a descending limb (D) permeable only to water -- a task performed by the aquaporins in the kidney --, while the second ascending limb (A) is coated with 'activator' pumps  -- accounting for the sodium pumps in the kidney --~\cite{Greger1996,Chou1999, Nielsen2002, Greger1981}; (iii) this coil is embedded in a common loose material, permeable to both water and salt, the interstitium; (iv) the first $U$-loop is continued by another loop and the so-called collecting duct (CD), again permeable to water only. {This model is inspired by the so-called central core models of the nephron~\cite{Jacquez1976,Foster1976,Stephenson1972,Step73-a,Step73-b,Step73-c,Step74,Step87} (see also Supplemental Material Sec.~I). In our model though, we do not wish to accomplish a faithful description of the kidney and simplify the process to its elementary ingredients. This will also allow us to get detailed insights in the separation mechanism.}

The initial solution entering the device from the D top is an aqueous solution with a waste to be extracted -- in the case of the kidney, urea--, with respective concentrations $\Wi$ and $\Si$. Aside from the specific geometry of the coil, an essential feature of the process is to use an `osmotic activator' -- in the case of the kidney, NaCl salt -- that enters the D limb along with the mix, with concentration $\Ai$. The terminology of 'osmotic activator' is justified by the role played by the salt in this separation process, and constitutes the main working principle of this loop: thanks to the $U$-loop geometry, the pumps in the A limb generate a salinity gradient in the interstitium which drags passively, via  osmosis, the water from the D limb to the interstitium. In simple words, the work performed by the ionic pumps is further harvested to also drag water osmotically in the interstitium. The geometry of the loop of Henle thus plays the role of an {\it osmotic exchanger}. 

\subsection{Transport equations and osmotic fluxes}
We now analyze quantitatively the transport of the various components along the serpentine geometry sketched in Fig.~\ref{fig1}. For the sake of simplicity, we develop a one dimensional modelization along the tube length. The coordinate along the tube is $x$, and the origin is located at the top of the D limb. The length of the tube from top to first loop is $L$ (see Fig.~\ref{fig2}b). This simplification does not alter the main ingredients. One may indeed check that the equilibration of the concentrations by diffusion processes in the orthogonal direction is fast compared to the axial velocity of the fluid, so that concentrations may be considered uniform in the orthogonal direction \sop{within each limb (but they may strongly differ from limb to limb)}. Moreover, the velocity of the fluid in the limbs is high enough so that diffusion processes along the axial direction can be also neglected. \sop{We consider for simplicity the steady state regime of the system.}

In the following we first focus on the single $U$-loop, \textit{e.g.} the Loop of Henle {\it per se} (D+A limbs), and extend then our analysis to the complete double loop geometry (D+A limbs +CD). 

\vskip0.2cm
{\it Descending limb --} Along the D limb, the water flux evolves due to the permeation of water under the osmotic pressure across the semi-permeable D limb walls. 
Writing the infinitesimal water balance and osmotic fluxes~\cite{Finkelstein1987} along a slice of the D limb allows to write the equation for the water flux along D as:
\begin{equation}
\label{eqWater}
 \dx{v_D\W_D} = \frac{2P_f}{r} \ln\left(\frac{a_I^W}{a_D^W}\right)
\end{equation}
where $v$ is the velocity of the fluid, $P_f$ the permeability of the D walls, $r$ the radius of the limb, $a^W$ the chemical activity of water -- here, assumed to be proportional to the molar fraction of water --, D (resp. I) indices refer to variables in the D (resp. the intersitium). In the D limb, all other species are conserved. 

The fluid velocity can be calculated thanks to conservation of mass in every limb. For instance in the D limb, this writes: 
$\dx{v_D} = \vW\left(\dt{\W_D} + \dx{v_D\W_D}\right)$, 
with $\vW$ the molar volume of water. The hydrostatic pressure drop along each limb is ruled by Poiseuille flow. However, the pressure drop is significantly small~\cite{Palatt1973}, and is thus neglected. Accordingly, we can safely neglect in Eq.~(\ref{eqWater}) the contribution of the hydrodynamic pressure drop to the osmotic pressure drop. 

\vskip0.2cm
{\it Ascending limb --}
In the A limb, the salt -- denoted in the following as the 'osmotic activator' -- is actively pumped across the walls.
We consider a general case where the pumps drive $n$ osmotic activator molecules for an elementary energy cost. In the kidney, 
this is provided by the dissociation energy of one adenosine triphosphate (ATP) molecule and the \sop{stoechiometry} is believed to be 3 $\mathrm{Na}^+$ for 1 ATP~\cite{Greger1983}; \sop{$Cl^{-}$ follows through the tissue walls and diffuses quickly to achieve electroneutrality}~\cite{Greger1983}. We write Michaelis-Menten kinetics to describe the activator concentration evolution in the A limb~\cite{Yang2014,Garay1973}:
\begin{equation}
\label{eqSalt}
 \dx{v_A\A_A} = - \frac{2nV_m}{r} \left( \frac{\A_A}{K+\A_A}\right)^n
\end{equation}
where $V_m$ is the maximum rate intake, $K$ is the Michaelis constant, and A indices refer to variables associated with the A limb. \sop{The kinetics involved are assumed to be independent of the concentration of osmotic activator in the interstitium, and furthermore, the energy requirement for the pump is assumed to be independent of the concentration of any of the constituents}~\cite{Greger1983}. The kinetics described by Eq.~(\ref{eqSalt}) can be applied to various pumping mechanisms. All other species are conserved in the A limb. \sop{In particular, the membrane separating the A limb and the interstitium is considered impermeable to water.}

\vskip0.2cm
{\it Interstititum --} In the Interstitium, equations similar to Eq.~(\ref{eqWater}) and~(\ref{eqSalt}) are written for the salt and water concentrations, with the sign reversed for the terms on the right hand side. No waste is present in the interstitium. \sop{The velocity of the fluid is taken to be zero at the bottom of the interstitium. }

\section{Results: spatial distributions, fluxes and optimal separation ability}

The previous analysis yields a set of 11 coupled and self-consistent non-linear transport equations for the water, osmotic activator and waste concentrations, as well as for the velocities in the various compartments. At the entrance of the device, the fluxes of water, osmotic activator (salt) and waste are prescribed as $\qWi$, $\qO$ and $\qS$. The integration of the previous system of equations yields the spatial distributions of the molar fractions of each component and the outcoming water and waste fluxes, $\qWo$ and $q_{Waste}^{\mathrm{out}}$.
The whole set of equations and boundary conditions is recalled in the Supplemental Material, Sec. II.

This complex system of differential equations is first solved numerically using standard methods, see Materials and Methods. We also perform below a systematic analysis of the transport equations at steady state. This provides several analytical results for the spatial dependence of the concentrations and fluxes of the various components, as well as reliable estimates of the separation ability and energy cost of the system. 

\begin{figure}[h!]
\includegraphics[width = 8.7 cm]{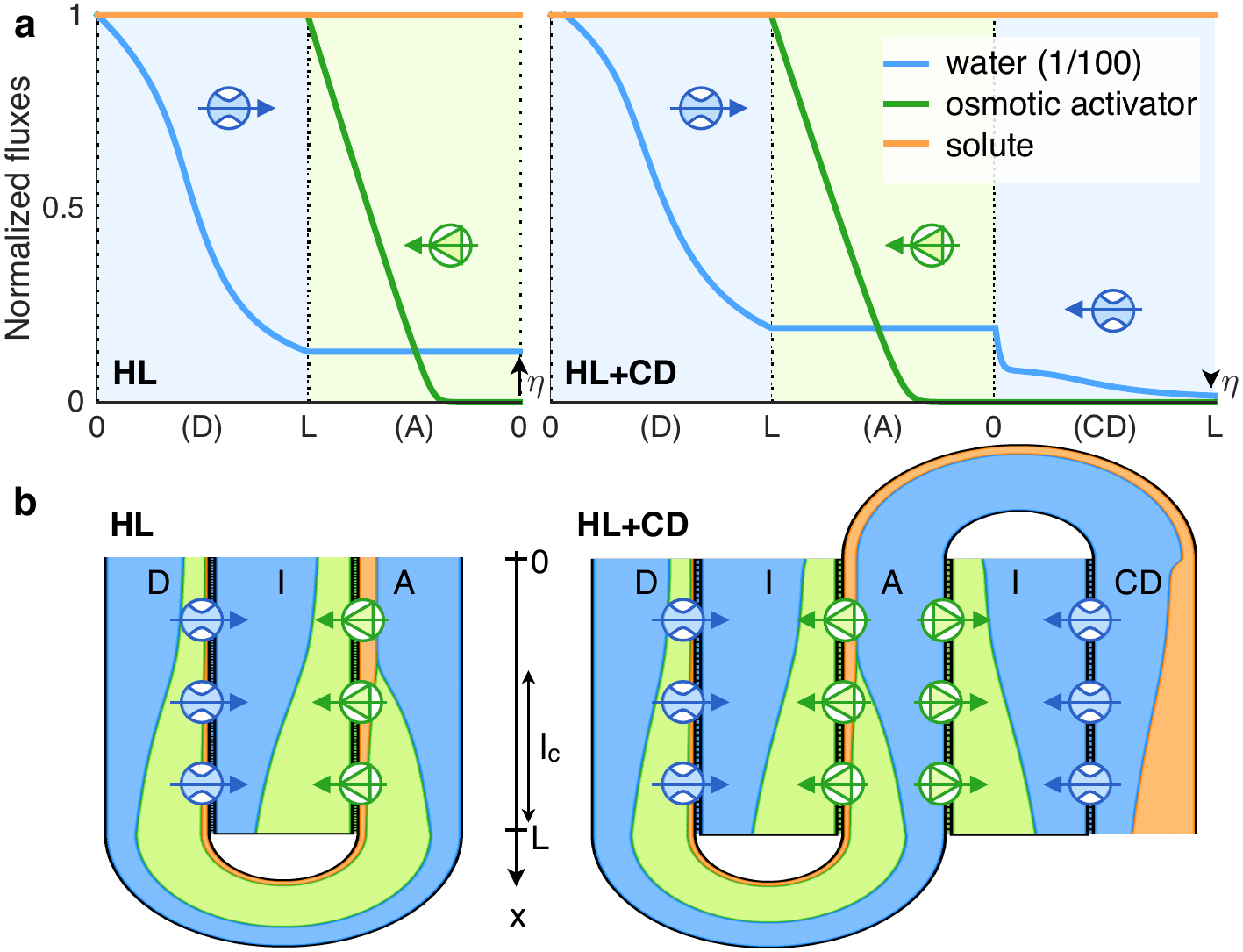}
\caption{{\bf Spatial distributions of the fluxes and molar fraction in the $U$ loop and full double loop.} (HL and HL+CD models) (a) Fluxes of water (divided by a scaling factor of 100), Osmotic activator (salt) and \sop{waste} along the descending (D) and ascending (A) limbs, left panel; and along the Collecting Duct (CD)   for  the HL+CD case; in this graph, the water loss ratio $\eta  = \qWo/\qWi$ is the final normalized flux.(b) Spatial distribution of the molar fractions of water (divided by a scaling factor of 100), Osmotic activator and Waste along the D and A limbs and in the interstitium. Note that in the (HL+CD) double loop geometry, the interstitium exchanges with all three branches (D,A,CD); although 2 were plotted for convenience, they have the same composition.  Numerical data are calculated with parameters $L=4.3 mm$, $P_f=2500 \mu m/s$, $n = 1$.}
\label{fig2}
\end{figure}

A typical numerical result is shown in Fig.~\ref{fig2} for the spatial evolution of the fluxes and molar fractions of the various components along the $U$-loop and double loop geometries, at steady state.

We explore now in details the results in order to gain some insight on the filtration efficiency.

\subsection{Osmotic activator pumping}

We start the analysis by studying the absorption of the osmotic activator in the ascending limb (A). As highlighted in
Fig.~\ref{fig2}, what emerges from the numerical calculations is the existence of a characteristic length for reabsorption.  
This is confirmed by the analytical resolution of the transport equation for $\A_A$ and $v_A$ in the A limb. 
Typically this length scale can be constructed by balancing the input flux of osmotic activator, $\qO=\pi r^2 v^{\rm in} \A^{\rm in}$ with
the outward pumped flux, $q_{\rm pump}=2\pi r\,\ell_c\, V_mn $ (with $V_m$ the intake rate of the pumps), so that
\begin{equation}
\label{eqlength}
\ell_c = \frac{\qO}{2\pi r V_mn},
\end{equation}
This result can be also obtained from the equations by approximating the spatial derivative of the activator flux $v_A\A_A$ in the A limb by its value at the bottom of the loop ($x=L$) in Eq.~(\ref{eqSalt}). This simplified expression for $\ell_c$ is then obtained under the assumption that  
$K \sim 30$ mmol/L $\ll \A^{\mathrm{in}} \simeq 100-200$ mmol/L (Supplemental Material  Sec.~III and~Fig.~S2). 
Using typical values for the various parameters entering this equation~(Supplemental Material Table~1), we find $l_c\simeq 1 mm$, which compares well to the total length of the nephron~\cite{Knep77,Koep72}. Beyond that length scale, the activator uptake is negligible. 

Altogether salt reabsorption occurs in a region of length $\ell_c$ close to the bottom of the $U$-loop. 
In the following this allows to split the effective domain of investigation into two different regions, where we can quantify every variable in the A limb: the deep domain, in a region of length $\ell_c$ close to the bottom, where osmotic activator/salt reabsorption happens on the length scale $l_c$; and the higher, remaining domain. 

\sop{A preliminary conclusion is that the length scale of salt reabsorption $\ell_c$ does not depend on the length of the limb $L$. As a result, even if $L$ increases, salt will always be reabsorbed on the same length scale $\ell_c$ at the bottom of the A limb. However, the region of length $L-\ell_c$ at the top of the interstitium increases. In this region, concentrated salt coming from the bottom of the intersitium continues to be diluted by progressive water uptake under the osmotic pressure} (see Fig.~\ref{fig2}b). 

\subsection{Water reabsorption and optimal separation ability}

With this result at hand, we can now turn to the detailed balance in water reabsorption. A key question is how much water may be extracted with this process. We analyze both the $U$-loop and the double-loop geometry.

Because the velocity at the bottom of the interstitium vanishes, the osmotic activator accumulates there, increasing the osmotic pressure between the D limb and the interstitium. As a result, water is dragged out of the D limb into the interstitium. The concentration of osmotic activator in the D limb thus increases, and water leakage is possible until the osmotic pressure between the D limb and the interstitium equilibrates. 

\vskip0.2cm
{\scshape Maximum separation ability in the $U$-loop  --}
Let us start with the investigation in the $U$-loop. Because the exchanges between the intersitium and the limbs are well quantified, one may express the variables of the interstitium according to the variables of the D limb (unknown so far) and the variables of the A limb (with known approximates). Accordingly, a straightforward derivation allows to obtain   two self-consistent sets of equations for the variables of the D limb alone, which can be solved although still complex. We report in the Supplemental Material the details of the analytical calculations.

\begin{figure}[h!]
\center
\includegraphics[width = 8 cm]{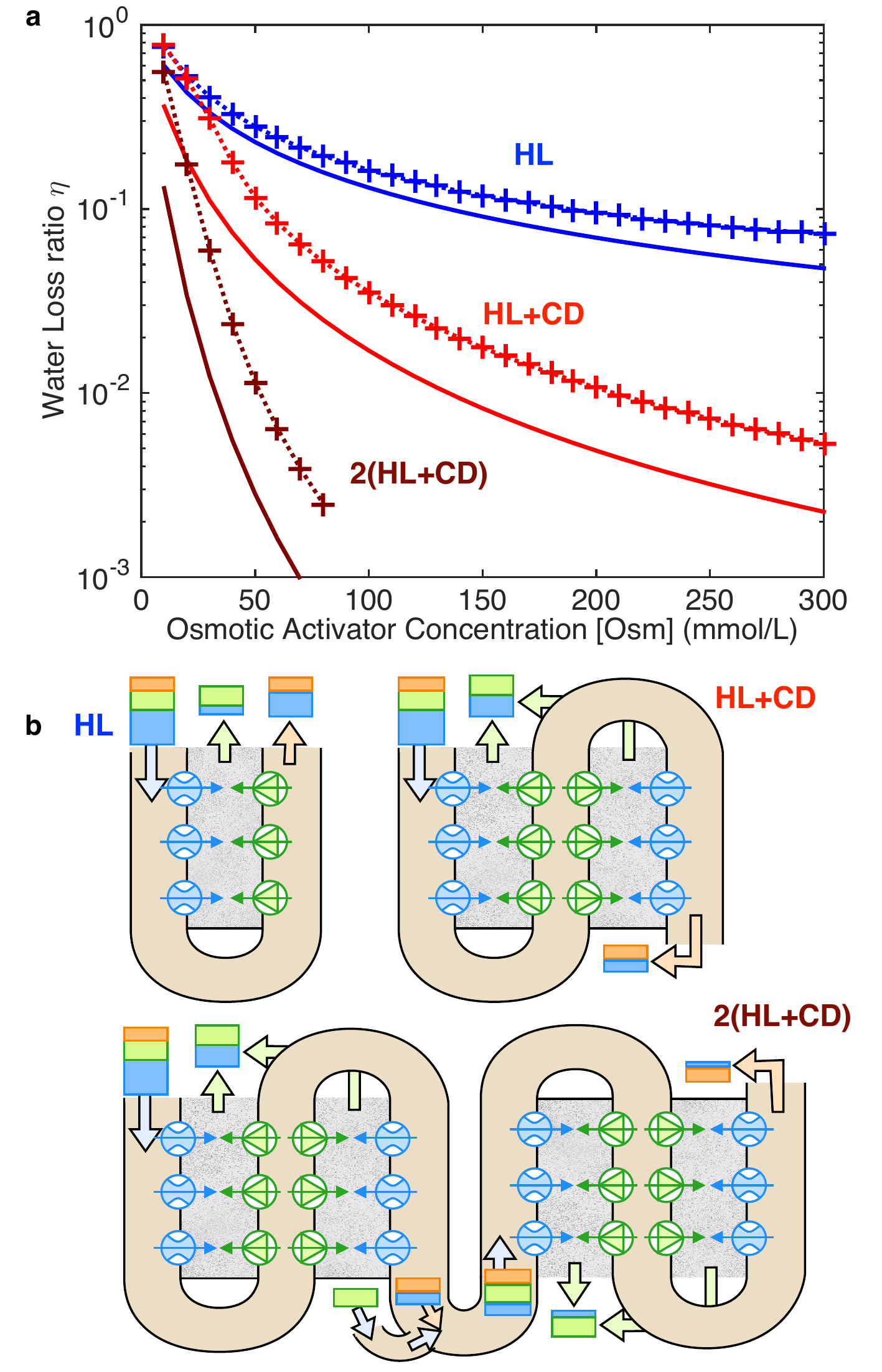}
\caption{\textbf{Separation ability and minimal bounds.} a - Water loss ratio $\eta  = \qWo/\qWi$ as a function of the initial concentration of osmotic activator $[Osm]^{\mathrm in}$ (salt). Both the numerical results (crosses and dotted lines) and the predicted lower bounds (solid lines) are reported, for the single loop (HL), double-loop (HL+CD) and two double-loop cycles (2 HL+CD). The latter quantities are given by Eqs.(\ref{eqEta}) and (\ref{eqEta2}) for the HL and HL+CD geometries, respectively. Numerical data are calculated with parameters $L=4 mm$, $P_f=2500 \mu m/s$, $n = 3$. b - Schematics of the nephron-inspired systems studied in a. }
\label{fig3new}
\end{figure}

Beyond the detailed solutions, some helpful analytical predictions may be extracted from these calculations. This concerns in particular  
the proportion $\eta$ of water flux remaining in the tube, {\it i.e.} the water loss ratio, $\eta=\qWo/\qWi$. In Fig.~\ref{fig3new},
we plot the numerical results for $\eta$ as a \sop{function} of the initial osmotic activator (or salt) concentration both in the $U$-loop,
and in the double-loop geometry discussed below.

A simple yet key result can be obtained for the water loss ratio, which characterizes the separation ability.
It is obtained from the analysis of Eq.~(\ref{eqWater}) in the upper part of the limb, by identifying that the flux of water through the semipermeable wall should always be directed from the D limb towards the interstitium (see Supplemental Material Sec.~IV for a detailed derivation). 
This rewrites as:
\begin{equation}
\label{eqEta}
\eta = \frac{\qWo}{\qWi} \geq \eta_{\rm min}=\frac{\S^{\mathrm{in}}}{\S^{\mathrm{in}}+\A^{\mathrm{in}}},
\end{equation}
where we recall that $\S^{\mathrm{in}}$ is the initial concentration of solute to be extracted arriving in the D limb and $\A^{\mathrm{in}}$ of osmotic activator. 

This lower bound can be interpreted physically in terms of the osmotic pressure balance between the D limb and the interstitium. Indeed for the water flux to be directed from D to I, the chemical balance requires that 
$a_I^W \le a_D^W$, see Eq.(\ref{eqWater}). This condition can be rewritten in terms of the various fluxes.
On the one hand the fluxes entering the D are $\qWi$ for water, $\qO$ for the osmotic activator, $\qS$ for waste. On the other hand, by conservation of mass, the fluxes exiting the interstitium are $\qWi - \qWo$ for water, $\qO$ for the osmotic activator {\it if it has completely been reabsorbed}, and $0$ for the waste. 
In this minimal situation, the previous osmotic pressure balance yields:
\begin{equation}
 \frac{\qWi - \qWo}{\qWi - \qWo + \qO}  \le \frac{\qWi}{\qWi + \qO + \qS}.  
\end{equation} 
and some simple algebra allows to recover Eq.~(\ref{eqEta}). 

	 
		
Note that it is also possible to derive an (approximate) upper analytic bound $\eta_{\rm max}$ for the water loss ratio $\eta$, by solving the equations on the species of the $D$ in the higher and deeper parts of the limb, see Supplemental Material Sec IV.C. The upper bound $\eta_{\rm max}$ converges to the lower bound $\eta_{\rm min}$ defined by Eq.~(\ref{eqEta}) in the limit of large pore permeability and/or long tube (Supplemental Material Fig.S3). 
For typical parameter values in the kidney, doubling the permeability decreases the water loss ratio by $20\%$ and doubling the length by $60\%$. The lower bound on $\eta$ can also be approached, {\it i.e.} $\eta_{\rm max} \rightarrow \eta_{\rm min}$, by allowing for a non-uniform spatial distribution of the water permeability along the D walls or of the pumps along the A walls (Supplemental Material Sec.~III-C), as observed in Nature. 
In the general case, the lower bound for $\eta$ in Eq. (\ref{eqEta}) provides a good approximation for the 
variational dependence of $\eta$ versus the initial osmotic activator (or salt) concentration,  as shown in  Fig.~\ref{fig3new}.

The minimal bound on the water loss ratio in Eq.~(\ref{eqEta}) is a key result because it provides a fundamental measure of the separation ability of the system. To achieve a good separation of waste from water in this device, the outflux of water $\qWo$, including the waste, should be as small as possible, {\it e.g.} that the water loss ratio $\eta$ be as small as possible. Eq.(\ref{eqEta}) shows that the uncovered water $\eta$ is limited by the initial molar ratio of the solute to be extracted to the one of the osmotic activator. So, no matter how efficient the activator pumps are or how high the water permeability of the membranes is, it is not possible to recycle more water from the system than $1 - \eta_{\rm min}$.

{Another way to interpret the result of Eq.~\ref{eqEta} is to think of the osmotic exchanger as a concentrating device for the waste,
{\it i.e.} any solute to be separated from water. 
 The concentrating ability of this system, namely $\So/\Si$, can accordingly be expressed as a function of $\eta$ as $\So/\Si = 1/\eta$. Therefore Eq.~\ref{eqEta} is also a measure of the maximal concentrating ability of the system.}

For typical parameter values in the kidney, the water loss ratio is $\eta \sim 0.2$ and is limited from below by $\eta_{\rm min}\approx 0.1$ after the first $U$ loop of Henle.  
At this point we note that such a minimal water loss ratio is still in the high range of physiological data~\cite{Greger1996}. For a human being, with $\eta \sim 0.1$ and an average flux of fluid (water, urea and salt) through the nephrons of about $120$ mL/min, the daily water loss would be tremendous, $\sim 18$ L, and accordingly not viable. This separation ability is also too low to be technologically relevant. 

\vskip0.5cm
{\scshape Maximum separation ability in the double-loop --}
Remarkably, a solution to bypass this limitation has already been achieved by Nature. 
One may indeed observe that, in addition to the $U$ geometry of the Henle loop, the nephron exhibits a second limb:  the so-called collecting duct (CD), that is only permeable to water as the D limb, see Figs.~\ref{fig1}-~\ref{fig3}. 

A key point is that this second limb is in contact with the same interstitium, therefore allowing to reabsorb water for a second time. 
From the theoretical point of view, the analysis follows the same lines as above (see Supplemental Material Sec.~VIII). \sop{The concentration of salt in the intersitium is still assumed to be well mixed or homogeneous in the orthogonal direction. Water is reabsorbed along the CD, following a permeability law similar to Eq.}~\ref{eqWater}\sop{, the key difference being that solute entering the CD has a different composition than the solute entering the D limb.} 
A modified bound for the separation ability of the double-loop geometry is now given by: 
\begin{equation}
\label{eqEta2}
\eta\geq \eta_{\rm min}^{HL+CD}=\left[{\S^{\rm in}\over\S^{\rm in}+\A^{\rm in}}\right]^2.
\end{equation} 
In Eq.(\ref{eqEta2}), the square dependence, as compared to Eq.~(\ref{eqEta}), originates from this second absorption step. As above, the lower bound can be reached with an adjustment of geometrical and physiological parameters of the loops. Typically one finds now that $\eta_{\rm min}^{HL+CD}\approx 0.01$, yielding a $\sim 2$ L daily water loss for a human being, very close to usual physiological observations~\cite{Greger1996}. This double loop device is therefore far more efficient than the simple Henle loop,  Fig.~\ref{fig3new}. 
As for the single $U$-loop, the lower bound for $\eta$ in Eq. (\ref{eqEta2}) provides a good approximation for the 
variational dependence of $\eta$ versus the initial osmotic activator (or salt) concentration, see Fig.~\ref{fig3new}.

Now, putting more than two loops in series (for instance additioning a third loop with a limb covered with osmotic activator pumps) does not improve further the performance because the osmotic activator uptake only happens in the first ascending limb. However, successive cycles in the HL+CD system can be done, provided that a certain amount of osmotic activator is additionned to the mix before each new cycle. In the case of two (respectively $N_c$) successive cycles, if the concentration of osmotic activator at the beginning of each HL+CD cycle is re-initiated, the minimal water loss ratio will be that defined by Eq.(\ref{eqEta2}) now squared (respectively to the power $N_c$). The improvement yielded for 2 cycles is plotted as well in Fig.~\ref{fig3new}. 

Eq.(\ref{eqEta2}) is the first main result of this paper. It summarizes in a compact formula how the double-loop geometry acts as an osmotic exchanger to efficiently concentrate a waste. A better waste concentration and higher water recycling ability is achieved when the osmotic activator/salt concentration is increased. Indeed, the larger the input salt concentration, the stronger the osmotic gradient in the interstitium that allows for an increased reabsorption of water. However, a higher salt concentration requires more energy to pump the salt from the A limb to the interstitium. \sop{This raises the question of the energetic performance of the HL and HL+CD systems, in particular if we want to consider them as working useful filtration devices.}

\section{Energetic performance of the osmotic exchanger}

Beyond the separation ability estimate of the device, the energetic performance of the process remains to be assessed. From a physiological point of view, it is indeed vital for the kidney to operate at a minimal energy cost in view of the amount of water processed every day. {Although some estimates of the free energy expense in models of nephron  have been developed, they all fail to account for the energy provided by the salt pumps~\cite{Newburgh43, Edwards, Layton2013, Kuhn2001, Step73-a, Hervy2003}}. Also, it is important to compare the energy expense of such an osmotic exchanger to more standard filtration devices.

To simplify, we assume that the permeability, pump speed and tube length are adjusted so that $\eta$ reaches its lower bound $\eta_{\rm min}$ in Eq.~(\ref{eqEta}) or Eq.~(\ref{eqEta2}), depending on the geometry, respectively $U$-loop or double-loop.

\subsection{Energetic cost for the single and double loop osmotic exchanger}

\begin{figure}[h!]
\begin{center}
\includegraphics[width =8.7 cm]{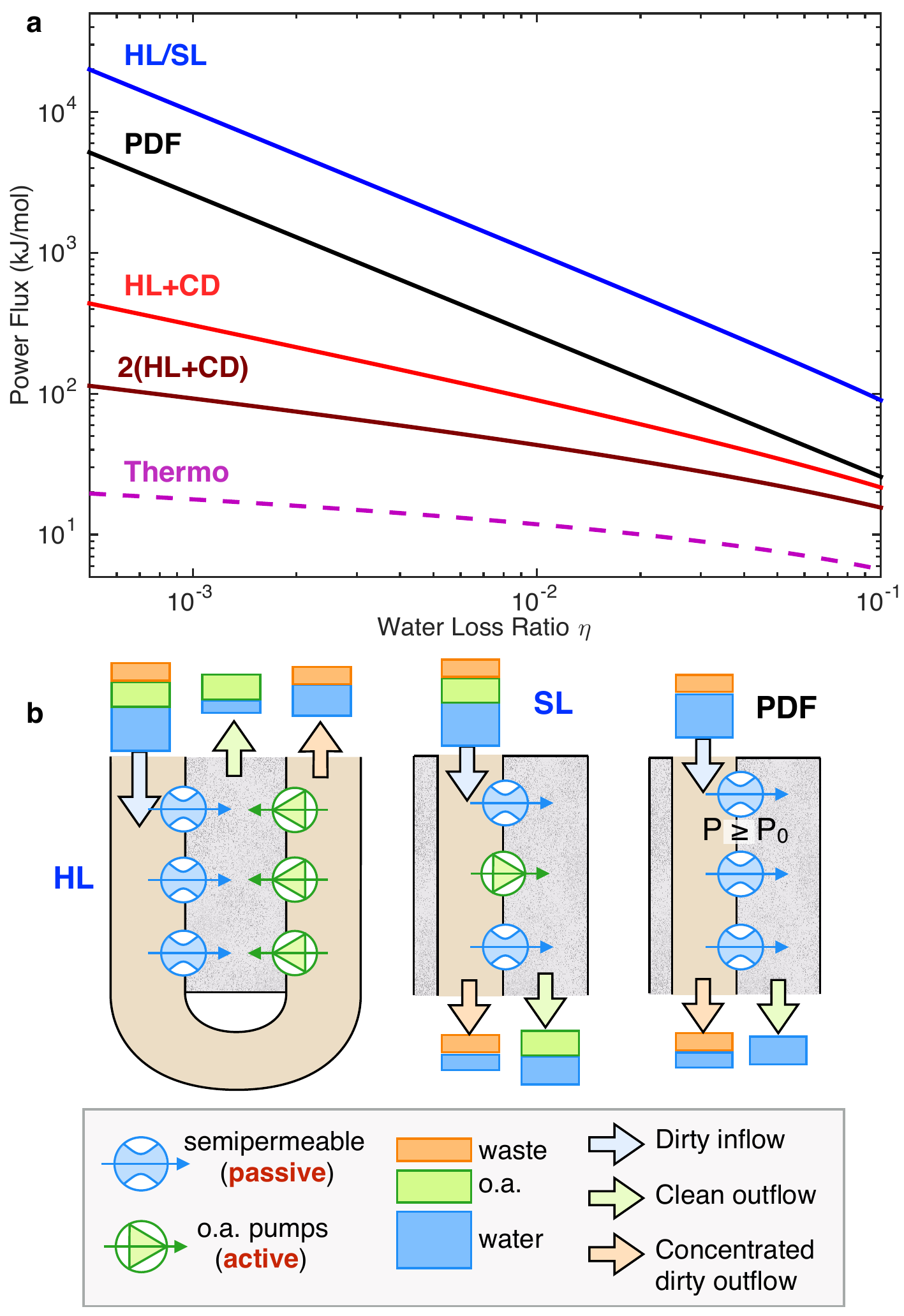}
\caption{\textbf{Comparison of filtration efficiencies of various systems.} Power required versus water loss ratio for the different systems. The data points are from the analytical expressions of the text. 2(HL+CD) corresponds to 2 successive HL+CD cycles, with addition of osmotic activator in between the two cycles. Thermo corresponds to the minimal energy required to achieve a separation defined by a given water loss ratio. The nephron working conditions correspond to  $\eta \approx 0.01$. Inset: sketch of corresponding systems: HL: single $U$ loop; HL+CD = double loop; SL = Single Limb; PDF = Pressure Driven Filtration. See text for detail. } 
\label{fig3}
\end{center}
\end{figure}

In the system described in Fig.~\ref{fig1}, the energy consumption reduces to the energy required for the pumping of the osmotic activator along the A limb.
The consumed power is written as 
\begin{equation}
\cP_{HL}=\int_0^L \pi r^2 dx {e_{ATP}\over n} N_{Osm}(x)
\end{equation} 
with $N_{Osm}(x)$ the pumped flux of osmotic activator across the A membrane, and $e_{ATP}$ the required energy to drive $n$ osmotic activator molecules within a single pump, which in the case of the kidney corresponds to that of 1 ATP molecule. Using mass conservation $N_{Osm} (x)=-{d\over dx}[ v_A(x)\A(x)]$ and $v_A(L) \A_A(L)=v_D(0)\A_D(0)$, one may integrate explicitly the previous equation to obtain
\begin{equation}
\label{eqPowerHL}
\displaystyle \cP_{HL} = \qO e_{ATP}/n \underset{\eta\rightarrow 0}{\simeq} \qS \frac{e_{ATP}/n}{\eta}.
\end{equation}
This is a simple and compact prediction showing that the power cost $\cP$ diverges like the inverse of the water loss ratio $\eta$. As expected it becomes increasingly costly to obtain a higher separation of the waste, {\it i.e.} $\eta \rightarrow 0$ (Fig.~\ref{fig3}b, HL).

This result describes the power consumption for the single loop exchanger.
A much better energetic efficiency is actually reached by the double-loop system, in direct line with its
improved separation ability.
Indeed, since the interstitium is common to the $U$-loop and the collecting duct, the same amount of activator pumping is harvested to reabsorb water both from the D limb and the CD. Thus, no further energetic cost is required as compared to the single $U$-loop, although for a higher separation. Collecting the results, one obtains:
\begin{equation}
\label{eqPowerHLCD}
\cP_{HL+CD} \underset{\eta\rightarrow 0}{\simeq} \qS\frac{e_{ATP}/n}{\sqrt{\eta}}.
\end{equation}
Eq.~(\ref{eqPowerHLCD}) demonstrates that the power consumption of the double-loop system is considerably reduced compared to the single $U$-loop (Fig.~\ref{fig3}\sop{a}).

In the case of two successive cycles in the HL+CD system, with the addition of the same amount of osmotic activator to the mix before each new cycle as considered above, the energy expense is doubled -- since the osmotic activator has to be readsorbed twice -- but the minimal water loss ratio is squared. A two cycle process yields a power cost $\cP_{2\times(HL+CD)}\simeq 2\qS (e_{ATP}/n)/\eta^{1/4}$,  optimizing further the energy efficiency of the process for small values of $\eta$, see Fig.~\ref{fig3}\sop{a}. A $N_c$-cycle process would accordingly operate at a power cost $\cP_{2\times(HL+CD)}\simeq N_c\qS (e_{ATP}/n)/\eta^{1/2N_c}.$

The above results can be compared to the minimal thermodynamical energy required for separation. The latter is estimated by computing the minimal work of separation for a given inward flux. After some algebra, see Supplemental Material Sec.~V, this can be expressed in the simple form
$\cP_{Thermo} \simeq -\qS RT \ln\left(\eta\right)$. As shown in Fig.~\ref{fig3}\sop{a}, this result is below the previous predictions
in Eq.(\ref{eqPowerHL}) and (\ref{eqPowerHLCD}), as it should. However increasing the number of cycles as discussed above allows to get closer to this lower energetic bound.

Eq.(\ref{eqPowerHLCD}) is the second main result of this paper.

\subsection{Power comparison to traditional sieving processes}
In the context of energy efficiency, it is of utmost interest to compare the previous exchanger device with other traditional filtration \sop{or concentration} systems. For this purpose, we first consider a nanofiltration (reverse-osmosis like) system to extract the solute. This corresponds to a geometry with a single limb with water permeable walls, and acts as \sop{a pressure driven filtration system (PDF)}, see Fig.~\ref{fig3}\sop{b}; no osmotic activator is required in this case. Water and waste only enter the limb at a high pressure with the same composition as in the dirty inflow in the nephron. In this situation, the hydrodynamic pressure drop is included in Eq.~(\ref{eqWater}) and we define similarly the water loss ratio $\eta$ as the ratio between the outward to inward flux of water through the limb \sop{(\textit{e.g.} $\eta$ is the ratio of water flux between the clean outflow and the dirty inflow)}. \sop{In this case, the high hydrostatic pressure allows to drive water out of the descending limb, against the osmotic pressure, and concentrates waste up to a certain degree (see Supplemental Material Sec.~VII for detail). If small components are also present in water - such as salt - and are also allowed to pass through the semi permeable membrane, the principle remains the same and the results are unchanged. The power required to achieve this process is proportional to the mechanical energy to drive that amount of flow with the corresponding hydrostatic pressure drop. The pressure drop may be further linked to the water loss ratio $\eta$.  Some straighforward calculations along this idea, as detailed in Supplemental Material Sec.~VII, allow to predict the  power required to achieve this process as}:
\begin{equation}
\label{eqPowerRO}
\cP_{PDF} \underset{\eta\rightarrow 0}{\simeq} \qS\frac{RT}{\eta}.
\end{equation}		
Interestingly, Eq.~(\ref{eqPowerRO}) is similar to Eq.~(\ref{eqPowerHL}) except for the thermal factor $RT$ replacing the ATP energy, $e_{ATP}/n$. Quantitatively $RT$ is of the order of $2.5$kJ/mol, smaller than $e_{ATP}/n \simeq 10$kJ/mol~\cite{Albe02}. The two behave similarly as a function of $\eta$, see Fig.~\ref{fig3} and  the HL system and the PDF system are thus only different by a multiplicative factor. 
A major difference though is that the PDF system requires to bypass the osmotic pressure $P>P_0\equiv  RT \S^{\rm in}$ associated with the separation of the waste. This pressure can reach a few atmospheres depending on the concentration of the solute to extract and such 'reverse-osmosis' like nanofiltration requires a high mechanical integreity of the material. 

As an alternative geometry, we also consider a 'single limb' system corresponding to a single limb where water permeable pores and activator pumps are intertwined on the wall of the limb (Fig.~\ref{fig3}a, SL). For this device we keep the osmotic activator in the solution. 
A similar analysis as for the single $U$ loop shows that the single limb system is basically equivalent to the HL, see Fig.~\ref{fig3}b and Supplemental Material Sec.~VI. 

Altogether, Fig.\ref{fig3} demonstrates that the double-loop exchanger, with one or more cycle, outperforms traditional filtration systems in terms of power efficiency, while working at small pressure.

\section*{Discussion}

These results show that with the double-loop geometry of the collecting duct in series with the Henle loop, Nature has evolved towards a most efficient geometry to filter out urea. From a technological point of view, it would be therefore highly inspiring to reproduce such a filtration device. This requires both the specific double-loop geometry highlighted above, and the use of an osmotic activator in the solution. We argue that all required ingredients are at our disposal to fabricate such an artificial nephron. Semi-permeable membranes will play the role of water permeable D and CD walls, replacing aquaporin-coated membranes. For salt used as an osmotic activator, the ion pump functionality can be mimicked using a stack of ion-selective membranes, with an electric field as the driving force, similar in that spirit to electrodialytic processes~\cite{Post2007}. Note that any ionic specie is in general a good candidate for osmotic activator, for it can be easily manipulated with electric fields -- \sop{although monovalent species should be used to begin with --} . Altogether modern microfabrication technologies developed for microfluidics would allow to directly mimick the set-up in Fig.~\ref{fig1} \cite{Sales2010,Jong2006}, with several devices possibly working in parallel (for more details, we suggest a detailed implementation of the device in the Supplemental Material Sec.~IX).

From the point of view of transport, such an artificial device can be described by the very same equations as above, Eqs.~\ref{eqWater}-~\ref{eqEta}. The pumping energy is replaced by the electric power required to displace the ions, which writes as $ e = \mathcal{F}\Delta\cV/2$ for a pumping under a voltage drop $\Delta\cV$ across the ion-selective membranes and replacing $e_{ATP}/n$. The power consumption thus takes an expression similar to Eq.(\ref{eqPowerHLCD}):
\begin{equation*}
\cP_{elec} \underset{\eta\rightarrow 0}{\simeq} \qS\frac{\mathcal{F}\Delta\cV/2}{\sqrt{\eta}}.
\end{equation*}
with now $\mathcal{F}\Delta\cV/2 \simeq 5 - 50 kJ/mol$ (using a value of $\Delta\cV \simeq 0.1-1 V$~\cite{Post2007}). 



Altogether, mimicking an osmotic exchanger would allow the development of novel filtration devices and dialytic systems. A 'kidney on a chip' constitutes a paradigm change as compared to actual dialytic systems involving standard sieving process, which are difficult to miniaturize~\cite{Fissel}: 
it allows for a compact design, with low energy consumption. Due to its low working pressure, it also allows to use soft materials in its design.

Such systems could also be used advantageously in the pre-treatment stage of standard desalination processes. Pretreatment to separate the basic contaminants from salty water \cite{Elimelech2011} typically costs more than $1 kJ/L$ in reverse-osmosis plants~\cite{Semiat2008,Fritzmann2007}. {Contrary to standard electrodialysis, one key point of the 'kidney on a chip' is that the extraction of ions is done from the brine}. Using recent progresses on ion selective membranes (e.g. monovalent selective)~\cite{Nagarale2006,Li2015,Karnik2015}, the 'kidney on a chip' geometry could effectively provide decontamination and dionization of targeted heavy ions and significantly diminish water hardness. This would reduce the corresponding energy cost and considerably improve the global energetic efficiency of the desalination process.

\sop{More inspiration could certainly be drawn from the kidney. For instance, the ionic pumps responsible for sodium reabsorption in the ascending limb are able to discriminate between sodium and potassium ions}~\cite{Greger1983}. \sop{So far, no one but Nature succeeded in making such a monovalent specific membrane.}


\begin{acknowledgments}
We are grateful to L. Bankir, D. Fouque, and A. Edwards for interesting discussions on the kidney. The authors are grateful for the fruitful comments of B. Rotenberg, A. Siria, J.-F. Joanny and R. Karnik. 
S.M. acknowledges funding from a J.P. Aguilar grant of the CFM foundation. L.B. acknowledges support from ERC-AG {\em Micromegas}. 
\end{acknowledgments}

\bibliography{Kidney}

\noindent See Supplemental Material at {\tt [URL will be inserted by publisher]} for details on analytical derivations and computations.

\end{document}